\documentclass[reprint, superscriptaddress, amsmath, amssymb, aps, showkeys]{revtex4-1}
\usepackage{graphicx}

\begin{document}

\title{Quantifying urban areas with multi-source data based on percolation theory}
\author{Wenpu Cao}
\affiliation{Institute of Remote Sensing and Geographical Information Systems, School of Earth and Space Sciences, Peking University, Beijing 100871, China}	
\author{Lei Dong}
\email{Corresponding author: arch.dongl@gmail.com}
\affiliation{Institute of Remote Sensing and Geographical Information Systems, School of Earth and Space Sciences, Peking University, Beijing 100871, China}	
\author{Lun Wu}
\affiliation{Institute of Remote Sensing and Geographical Information Systems, School of Earth and Space Sciences, Peking University, Beijing 100871, China}	
\author{Yu Liu}
\affiliation{Institute of Remote Sensing and Geographical Information Systems, School of Earth and Space Sciences, Peking University, Beijing 100871, China}
\date{\today}

\begin{abstract}
Quantifying urban areas is crucial for addressing associated urban issues such as environmental and sustainable problems. Remote sensing data, especially the nighttime light images, have been widely used to delineate urbanized areas across the world. Meanwhile, some emerging urban data, such as volunteered geographical information (e.g., OpenStreetMap) and social sensing data (e.g., mobile phone and social media), have also shown great potential in revealing urban boundaries and dynamics. However, consistent and robust methods to quantify urban areas from these multi-source data have remained elusive. Here, we propose a percolation-based method to extract urban areas from these multi-source urban data. We derive the optimal urban/non-urban threshold by considering the critical nature of urban systems with the support of the percolation theory. Furthermore, we apply the method with three open-source datasets - population, road, and nighttime light - to 28 countries. We show that the proposed method captures the similar urban characteristics in terms of urban areas from multi-source data, and Zipf's law holds well in most countries. The accuracy of the derived urban areas by different datasets has been validated with the Landsat-based reference data in 10 cities, and the accuracy can be further improved through data fusion ($\kappa=0.69-0.85$, mean $\kappa=0.78$). Our study not only provides an efficient method to quantify urban areas with open-source data, but also deepens the understanding of urban systems and sheds some light on multi-source data fusion in geographical fields.
\end{abstract}


\maketitle
 	
\section*{Introduction}
How to define an urban area, the basic spatial unit for urban planning and studies, has been a long-standing problem for researchers and policymakers \citep{batty2006rank, gabaix2004evolution, rozenfeld2008laws}. This problem has become more important in recent decades because of the emergence of a large number of fast-urbanizing regions around the world (e.g., China and India). However, due to the complexity of the urban system, especially the fuzzy urban-rural transition, consistent and robust measurement to quantify urban areas has remained elusive.

For a long time, governments have relied heavily on the administrative boundaries to address urban issues (e.g., environmental and sustainable problems); and many location-based policies are also implemented based on the administrative divisions. However, administrative divisions are mainly divided by historical, political, and geographical reasons, making it difficult to reflect the socio-economic dynamics of cities. Additionally, administrative divisions are incomparable across different countries and periods \citep{wu2018zipf}. Therefore, some countries turn to employ socio-economic indicators (e.g., population, economic activity, and commuting) to re-divide urban areas. For example, metropolitan areas (MAs), the most commonly used socio-economic boundaries, define urban areas as closely related regions in terms of socio-economic connection \citep{berry1969metropolitan}. However, the construction of MAs has three main shortcomings. First, the detailed data (e.g., census data and commuting survey data) to construct MAs are lacking in many developing countries. Second, the data collection process for MAs is time-consuming and expensive, making it unable to capture the rapid urbanization process in fast-growing regions. Third, the standard to define MAs varies among countries. There is still a lack of a unified approach to obtain functional urban areas, which can be applicable to all countries.

Remote sensing data, especially the satellite-based data, provide continuous and consistent observations of urban activities on earth \citep{zhu2019understanding}. They are easily accessible for most countries and have been widely used to study urban dynamics at different spatial scales \citep{aubrecht2016consistent, mertes2015detecting, small2011spatial, taubenbock2019new, zhou2018global}. Based on different urban characteristics (e.g., multispectral information, light emissions, and morphological structures), several global urban maps have been derived from the remote sensing data, such as MODIS500m \citep{schneider2009new, schneider2010mapping}, GHSL \citep{corbane2019automated}, GlobeLand30 \citep{chen2015global}, and GUF \citep{esch2012tandem-x}. Meanwhile, some emerging urban data with humans as sensors, such as volunteered geographical information (VGI, e.g., OpenStreetMap) \citep{goodchild2007citizens} and social sensing data (e.g., mobile phone and social media) \citep{liu2015social}, have also shown great potential in revealing the socio-economic boundaries of cities \citep{jiang2015head, jiang2011zipf, long2016mapping}. In addition to these multi-source datasets, some new methods have also been developed to delimit urban areas \citep{cao2009svm, jiang2015head, liu2018high-resolution, long2016mapping, rozenfeld2008laws, trianni2015scaling, zhou2014cluster-based}. Especially, City Clustering Algorithm (CCA), which defined cities as the maximally connected populated areas, has attracted great attention due to its simplicity and efficiency \citep{jiang2011zipf, rozenfeld2008laws, rozenfeld2011area, vogel2018detecting}. Benefited from advanced computing techniques and easily accessed data sources, CCA or other data-driven methods can derive urban areas in a timely and simple manner. However, due to the complexity of the urban system and the fuzzy urban-rural transition, these methods still have difficulties in finding the optimal threshold that differentiates between urban and non-urban areas. Additionally, most of the previous studies use only one variable (e.g., population \citep{rozenfeld2008laws, rozenfeld2011area}, road networks \citep{jiang2011zipf, long2016mapping}, nighttime light emissions \citep{imhoff1997technique, zhou2018global}, or built-up areas \citep{huang2016mapping, schneider2010mapping}) to quantify urban areas. While human activities are coupled together, it is largely unclear whether different urban data could reflect the similar urbanization process. Therefore, it should be particularly helpful to develop a universal method that can objectively find the optimal threshold to delimit urban areas via multi-source data.

Complexity science of cities sheds some light on the optimal threshold problem. Urban systems, as typical self-organized systems, display some universal macroscopic patterns, such as Zipf's law \citep{krugman1996self, zipf1949human}, scaling laws \citep{bettencourt2007growth}, and fractal characteristics \citep{batty1994fractal}. Previous studies have shown that these macroscopic patterns emerge at the critical point of the urban system \citep{newman2005power}, and several physical models have been adopted to study the critical phenomena of cities \citep{goh2016complexity, makse1995modelling, makse1998modeling}. Notably, the percolation model, a typical model for studying complex systems \citep{christensen2005complexity}, was used on the road network data of Britain to show that the urban system emerges at the critical point of the percolation process \citep{arcaute2016cities, molinero2017angular}. These works inspire us to address the optimal threshold problem with the percolation model.

In this paper, we propose a novel method to extract urban areas from multi-source urban data. We adopt a broader definition of urban areas as maximally connected areas that have more urban elements (i.e., population, infrastructure, economic activity) than non-urban areas, and these three urban elements are widely acknowledged in the urban geography and urban economics fields to measure the urbanization process \citep{arcaute2016cities, rozenfeld2008laws, vogel2018detecting}. We find the optimal urban/non-urban threshold solely through the input data themselves by considering the critical nature of urban systems. Specifically, we traverse all potential thresholds and aggregate the urban units into a cluster system under each threshold. Based on the percolation theory, we get the optimal urban areas when the whole system is at the critical point. To verify our method, we investigate the geographical layouts of urban areas derived by three datasets (population, road, and nighttime light). Despite the datasets of great difference, we find that: i) our method can capture the similar geographical distributions of urban areas; and ii) the rank-size distribution of urban areas fits well with Zipf's law, a fundamental law of urban systems. We further validate our results in 10 cities of the world, using urban reference data based on Landsat imagery. The derived urban areas by different datasets show good agreement with the reference data, and the accuracy can be further improved through data fusion. These findings demonstrate the effectiveness of our method and also deepen our understanding of cities. From the perspective of applications, the efficient, consistent, and low-cost properties of this method make it a good starting point for mapping urban areas around the world.

\section*{Study areas and data}
\subsection*{Study areas}
We choose China as the main study country to demonstrate the effectiveness of our method. Then we apply our method to 28 countries to validate the universality of this approach. Despite the rapid urbanization experienced in the past few decades, many areas of China are still underdeveloped, and the rural-to-urban process is quite uneven across regions. More importantly, as the largest developing country, China is lacking in consistent urban area data, which highlights the meaningfulness of this study. For the remaining countries, we choose those largest countries in each continent, with an area of not less than $100,000km^2$, including both developed and developing countries (Table \ref{tab:country&data}). The national surface area information is derived from the 2016 United Nations Demographic Yearbook, available through United Nations Statistics Division. The administrative boundary data of all countries are available from GADM (\url{https://gadm.org}), an open-source database of global administrative areas.

\begin{table}[htbp]
	\caption{Study countries and urban data}
	\begin{ruledtabular}	
		\begin{tabular}{p{0.16\linewidth}p{0.2\linewidth}p{0.6\linewidth}}
			Study countries & Africa & Algeria, Chad, D. R. Congo, Libya, Sudan \\ \cline{2-3}
	 		& Asia & China, India, Indonesia, Kazakhstan, Saudi Arabia \\ \cline{2-3}
	 		& Europe & France, Germany, Spain, Sweden, Ukraine \\ \cline{2-3}
	 		& North America & Canada, Honduras, Mexico, Nicaragua, United States \\ \cline{2-3}
	 		& Oceania & Australia, Papua New Guinea, New Zealand \\ \cline{2-3}
	 		& South America & Argentina, Bolivia, Brazil, Colombia, Peru \\
			\colrule
			Urban data & Nighttime light & Global NPP-VIIRS nighttime light dataset \\ \cline{2-3}
	 		& Population & China mobile phone estimated population dataset \\
	 		& & Global population distribution dataset from WorldPop \\ \cline{2-3}
	 		& Road & China road network dataset from the Ordnance Survey \\
	 		& & Global road shapefile dataset from OpenStreetMap \\
		\end{tabular}
	\end{ruledtabular}
	\label{tab:country&data}
\end{table}

\subsection*{Data}
We use three datasets -- nighttime light (remote sensing data), population (social sensing data), and road networks (VGI data) -- in this research. These datasets represent the three most important urban elements: economic activity, population, and infrastructure, respectively.

\paragraph{Nighttime light data.}
We use the new generation of nighttime light (NTL) data, the global NPP-VIIRS NTL data (available through \url{https://www.ngdc.noaa.gov/eog/viirs}). The NPP-VIIRS NTL data is produced from the Visible Infrared Imaging Radiometer Suite (VIIRS) Day/Night Band (DNB). Compared with the old DMSP/OLS data, the NPP-VIIRS data has a higher spatial resolution (15 arcsec) and partially relieves the saturation effects and blooming effects \citep{shi2014evaluating}. Before averaging the observations, the annual composites exclude data impacted by stray light, lightning, lunar illumination, and cloud-cover. We collect the annual `vcm-orm-ntl' average radiance data for 2016, which has undergone the outlier removal process, with non-light background set to zero. The NTL data is publicly accessible for most countries, which make it a useful data source to map socio-economic activities and functional urban areas \citep{imhoff1997technique, small2011spatial, vogel2018detecting, zhou2018global}

\paragraph{Population data.}
We use two population data sources. The first one is the WorldPop dataset (available through \url{https://www.worldpop.org}). WorldPop provides an open-access archive of high-resolution population distribution data \citep{tatem2017worldpop}. Especially, the `Global per country 2000-2020' datasets have been improved in terms of global consistency. We collect the 2016 data for all study countries. For China, we also collect the second population dataset, which is estimated by the anonymous mobile phone location data. Detailed information about this dataset is presented in \cite{dong2017measuring}. Here, we use the aggregated version with a resolution of $0.001^\circ \times 0.001^\circ$. Note that mobile phone estimated population is only a sample of the whole population; thus, we scale up the data with a factor derived by (national population) / (number of mobile phone users in the sample).

\paragraph{Road networks.}
We collect the OpenStreetMap (OSM) road shapefiles (available through \url{https://download.geofabrik.de}) for all study countries. For China, we also collect the road network data from the Ordnance Survey, a more detailed dataset than the OSM data \citep{liu2016automated}. The raw ordnance survey data records every segment in road networks with two endpoints' IDs and locations. We identify road intersections by endpoint's ID and obtain about 21 million ones.

\subsection*{Data quality assessment}
For the nighttime light, the VIIRS data has been greatly improved with in-flight calibration, finer quantization, and lower light detection \citep{elvidge2013why}. Moreover, the annual `vcm-orm-ntl' dataset is obtained through massive cloud-free observations and eliminates the background noise and ephemeral lights, thereby enhancing the radiance stability across the world \citep{elvidge2017viirs}. Besides, our method relies only on the relative brightness values of different areas within a country. For the population, the Worldpop dataset has been proved to have high accuracy of population distribution as shown in \cite{stevens2015disaggregating}. The mobile phone estimated dataset comes from our previous work and also has high accuracy in measuring population distribution in China \citep{dong2017measuring}. For example, at the district (county) level, the $R^2$s of the regression between mobile phone inferred population and census population are 0.97 and 0.98 for Beijing and Shanghai, respectively. For the road data, the quality of OSM data varies greatly across countries regarding completeness and accuracy. However, due to lacking the `ground truth' of road data, we did not investigate the effect of missing data of OSM in this paper. As demonstrated by previous studies, OSM data can be a reliable data source for the task of mapping urban areas \citep{haklay2010how, jiang2011zipf}.

\section*{Methods}
\begin{figure*}[htbp]
	\centering
	\includegraphics[width=0.9\linewidth]{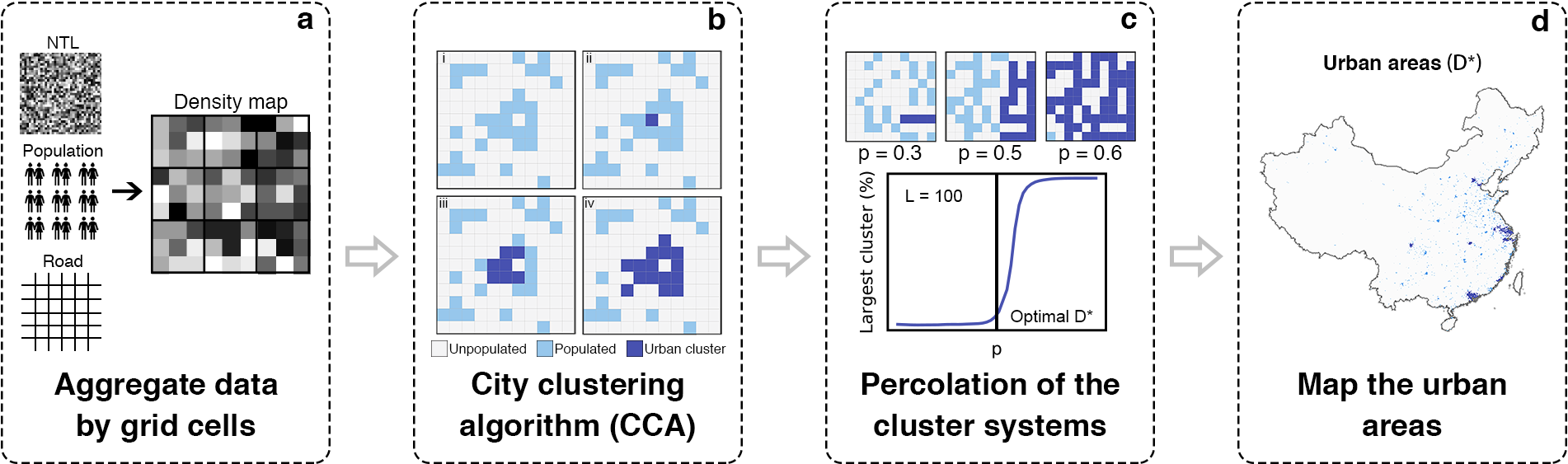}
	\caption{Schematic of the percolation-based city clustering method. (a) Multi-source urban data are aggregated by grid cells. 	(b) We use the CCA to merge urban units into a cluster system under each potential threshold. CCA: i) Cells with a greater value than the potential threshold are marked as urban units (light blue). ii) An unprocessed urban cell is selected to form a new urban cluster. iii) This urban cluster recursively adds the nearest urban cells until all nearest neighbors have been processed. We use the eight nearest neighbors in our research. iv) An urban cluster is formed (dark blue). Then another process begins until all urban cells belong to an urban cluster. (c) Two-dimensional site percolation model. As for a $L\times L$ lattice, each site can be occupied with probability $p$, and adjacent occupied sites form a cluster (light blue). As $p$ increases, the largest cluster (dark blue) remains stable at first, but quickly becomes a giant spanning one at the critical point. We regard each potential threshold as probability $p$ in percolation and find the optimal threshold $D^*$ at this critical point. (d) Urban areas with optimal threshold $D^*$.}
	\label{fig:PCCA_Schematic}
\end{figure*}

Our percolation-based city clustering algorithm (PCCA) includes three main steps. First, we aggregate the fine-scale urban data by $0.5^\prime \times 0.5^\prime$ grid cells to unify different datasets. Second, we apply the CCA to merge grid cells into urban clusters under each potential threshold. Third, we perform the percolation analysis on the detected clusters to find the optimal threshold and then map the urban areas at the optimal threshold. Fig. \ref{fig:PCCA_Schematic} shows the schematic of the PCCA. All steps will be discussed in detail below.

\subsection*{Aggregating multi-source data by grid cells}
Since multi-source urban data differ in granularity, data preprocessing is required to make the results comparable. For the nighttime light and population datasets, we directly downsample the data into $0.5^\prime \times 0.5^\prime$ grids. For the road datasets, we need more processing as raw data are vector files. For the Chinese ordnance survey data, we count the number of road intersections located in each grid cell and thus derive the road intersection grids. For the OSM data, to speed up the calculation, we divide road lines into small segments and count the total length of road networks in each grid cell. Then we obtain the road length grids. Note that we further divide the cell values of the population and road data by cell's spherical area to derive consistent density maps.

\subsection*{City Clustering Algorithm (CCA)}
At the grid cell level, we set each value of the above-mentioned datasets as a potential urban density threshold, which refers to the minimum density of population, infrastructure, or activity of urban areas. Cells with a greater value (more urban elements) than the threshold will be marked as urban units. Then, we apply the CCA to aggregate urban units into urban clusters under each potential threshold. The CCA originally uses fine-grained grid data of population and defines an urban cluster as the maximal, geographical continuous, populated areas \citep{rozenfeld2008laws}. Here, we expand the CCA not only to population, but also to all kinds of urban data (e.g., nighttime light and road networks). As shown in Fig. \ref{fig:PCCA_Schematic}b, an urban cluster starts with a random unprocessed urban cell, and recursively adds the nearest urban cells until all nearest neighbors have been processed. Then an urban cluster is formed. Another unprocessed urban cell is selected to form a new urban cluster until all urban cells belong to a specific cluster. We use the eight nearest neighbors in our research. To assess the stability of the approach, we also test the four nearest neighbors. A similar percolation process can be observed, and the optimal threshold and final urban maps remain stable, see Fig. S1. After performing the CCA, we get all cluster systems under each potential threshold.

\subsection*{Percolation of the cluster systems}
To find the optimal threshold of the CCA, we apply the percolation theory to analyze the properties of the extracted clusters. The percolation theory was originally developed in statistical physics and mathematics to study the emergent structures of clusters on a random graph. Since percolation can lead to some critical phenomena, urban researchers have then used percolation theory to model urban growth and to understand the critical phenomena of cities \citep{makse1995modelling, makse1998modeling, rozenfeld2008laws}. The two-dimensional site percolation is a simple and intuitive model to explain the percolation theory and explore the critical phenomena (Fig. \ref{fig:PCCA_Schematic}c). As for a $L\times L$ lattice, each site can be occupied with probability $p$, and adjacent occupied sites form a cluster. When $p$ is small, there are only a few small clusters. As $p$ becomes larger, the size of the largest cluster remains stable, despite more occupied sites. When $p$ reaches a certain point, a giant cluster quickly forms and spans the whole lattice. This point is called the critical point or the continuous phase transition. Around the critical point, the cluster system exhibits some critical phenomena (e.g., the size distribution follows power-law), which are also found in urban systems. Therefore, analogous to the two-dimensional lattice, we regard each potential threshold as the occupation probability $p$ in percolation, the optimal threshold $D^*$ can be found when the largest cluster of each cluster system becomes a giant one with a continuous phase transition. We consider the threshold at this critical point as the optimal threshold. After determining the optimal threshold, we obtain the final results of urban areas.

\section*{Results}
\subsection*{Urban areas extracted by PCCA}
We first apply the PCCA to the datasets of China. Following the Methods section, we obtain the density maps of population, road intersections, and nighttime light of China. Then, we extract the cluster systems under all potential thresholds and apply the percolation analysis to the cluster systems to find the optimal threshold.

\begin{figure*}[htbp]
	\centering
	\includegraphics[width=0.85\linewidth]{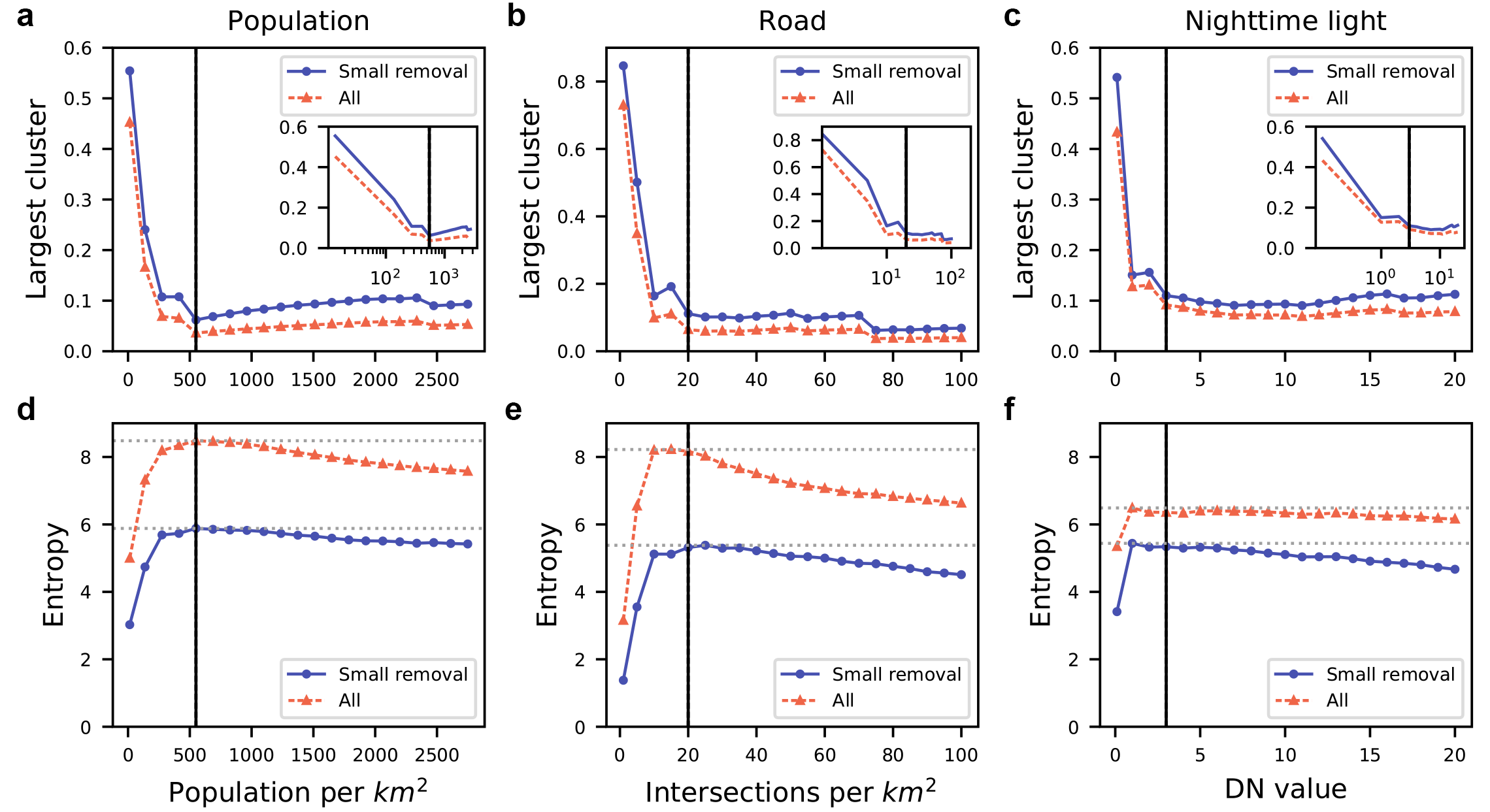}
	\caption{The normalized size of the largest cluster for different thresholds in the datasets of population (a), road (b), and nighttime light (c), and the distribution entropy of the cluster system for different thresholds in the datasets of population (d), road (e), and nighttime light (f). The solid blue lines and dotted red lines represent the results of clusters larger than $20 km^2$ and all clusters, respectively. The critical points of the continuous phase transition are marked with solid vertical lines. The values of these points are 550 people per $km^2$ in population data, 20 intersections per $km^2$ in road data, 3.0 DN value in nighttime light data. Maximum entropies are marked with dotted horizontal lines, which are also around the critical points. Insets: Similar to the main figures but with a log-scale for x-axis.}
	\label{fig:pcca_China}
\end{figure*}

Fig. \ref{fig:pcca_China}a-c presents the size of the largest cluster of different thresholds, and the size has been normalized by the total area of all clusters. At each threshold, there are some fragmented areas, possibly due to data noise or some special land use (e.g., oil fields, scenic areas). We set a minimum size -- $20 km^2$ -- to filter those fragmented areas. This value is set because the smallest land area of a city in China is approximately $20 km^2$. We also test the sensitivity of our method to this parameter by setting the minimum size to $10, 15, 25 km^2$, and the results are robust (Fig. S2). In Fig. \ref{fig:pcca_China}, solid blue lines present the results of clusters larger than $20 km^2$, and dotted red lines show the results of all clusters. For all datasets, as we lower the threshold, a giant spanning cluster quickly forms when the threshold reaches a critical point (vertical lines) -- indicating a continuous phase transition. This phenomenon reflects the characteristics of the urban system as an interconnected complex system. Since the intra-city connections are much stronger than the inter-city connections, weak inter-city connections break up as we increase the threshold. When the threshold reaches a certain point, all weak inter-city connections do not exist, while the intra-city connections can still be tied closely. As a result, the size of the largest cluster goes through a critical point, which we consider as the optimal threshold to quantify urban areas.

Besides the largest cluster, we also calculate Shannon's entropy $H$ of the size distribution for each cluster system: 
\begin{equation}
	H=-\sum_{i=1}^{N}{p_{i}\log{p_{i}}}
\end{equation} 
where $N$ is the number of clusters in the system, $p_i$ is the proportion of the area of cluster $i$ in all clusters. In Fig. \ref{fig:pcca_China}d-f, we find that the entropy also reaches the maximum (horizontal lines) around the critical point (vertical lines). Moreover, for each dataset, the entropies at the critical point are close for the clusters larger than $20 km^2$, which are 5.88 (population), 5.32 (road), and 5.34 (nighttime light), indicating the similar size distributions of urban areas extracted from different data sources.

\begin{figure*}[htbp]
	\centering
	\includegraphics[width=0.85\linewidth]{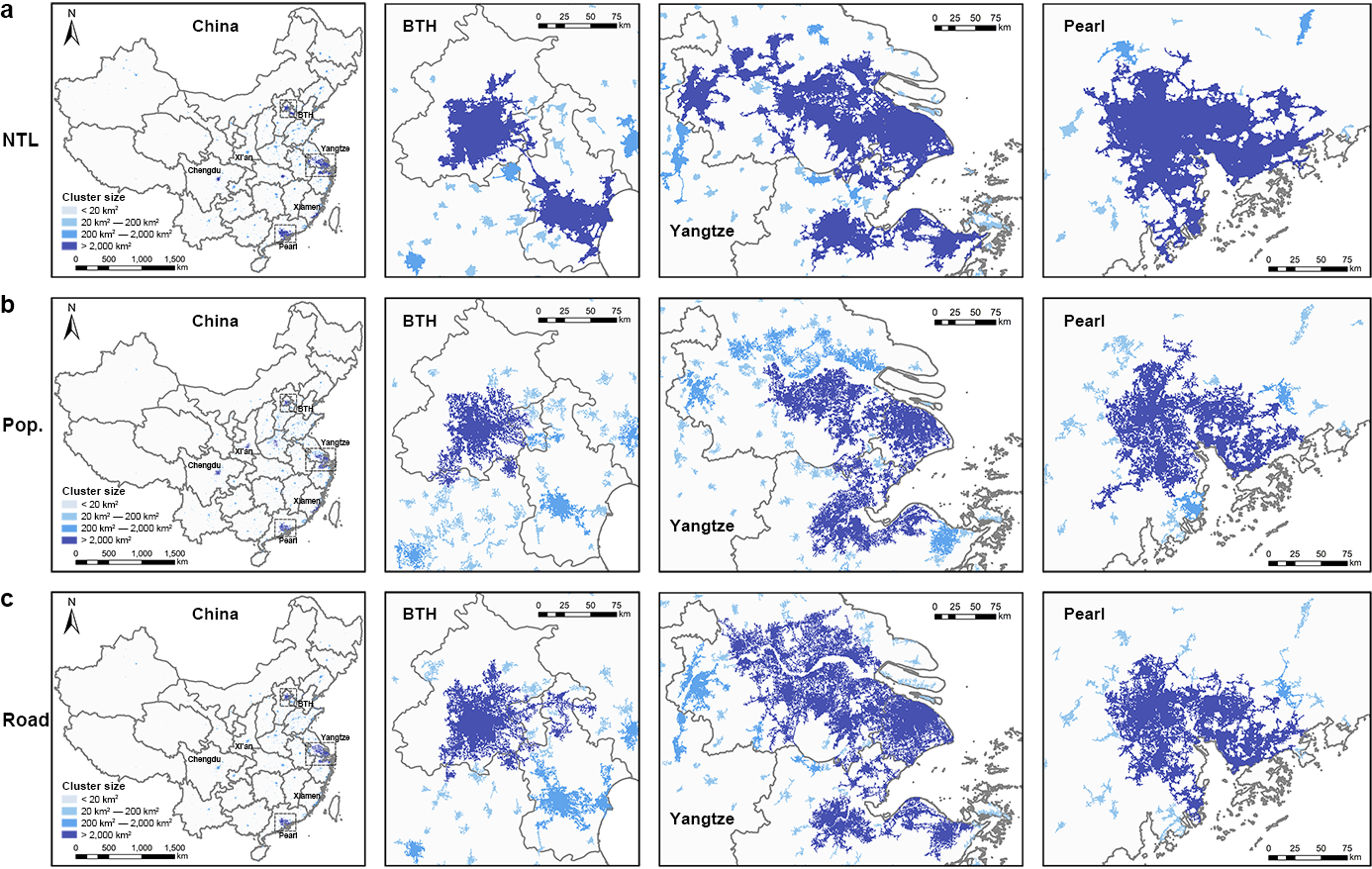}
	\caption{Urban areas in China delimited by the PCCA in the datasets of nighttime light (a), population (b), and road (c). The numbers of clusters ($> 20 km^2$) are 1085, 1260, and 931 for the nighttime light, population, and road datasets, respectively. Urban clusters are colored according to their geographical areas. Note that for simplicity, the South China Sea Islands are not shown in the maps.}
	\label{fig:natural-cities_China}
\end{figure*}

Furthermore, we map the urban areas at the critical point (Fig. \ref{fig:natural-cities_China}). Strikingly, different data yield similar results. Especially, those larger clusters are well-developed cities, such as Chengdu, Xi'an, and Xiamen. Moreover, the maps also echo the uneven regional development in China. The urbanization level is much higher in the southern and eastern regions, and the coastal areas are more developed than the inland areas. The top three largest areas delineated by our method are the Beijing-Tianjin-Hebei, the Yangtze River Delta, and the Pearl River Delta economic zones (enlarged view of Fig. \ref{fig:natural-cities_China}), which are the most urbanized megalopolis areas in China \citep{xie2016updating, liu2010spatial}. These areas break the geographic constraints of administrative boundaries and have highly integrated connections. Our results confirm their high integrations.

To measure the similarities of urban areas delineated by different datasets, we use the Dice similarity coefficient (DSC). DSC is a similarity measure over sets and ranges from 1, with the same sets, to 0, with two completely different sets. It is defined as:
\begin{equation}
	DSC=\frac{2\vert X \cap Y \vert}{\vert X \vert + \vert Y \vert}
\end{equation}
where $X$ and $Y$ are the sets of grid cells of urban areas. We remove the clusters smaller than $20 km^2$. The DSCs are 0.62 between population and road, 0.68 between road and nighttime light, and 0.59 between population and nighttime light, indicating that the spatial distributions of the urban areas obtained by different datasets are similar. Besides, we find the differences, those grid cells marked as urban areas in only one dataset, are mainly from the different distributions of intra-city `holes' and the peripheries of each urban cluster. For example, the Olympic Park in Beijing, with few people but dense roads, is a `hole' (non-urban areas) in the population map but urbanized in the road map.

We also investigate whether Zipf's law, one important law for the size distribution of the urban system, holds for our definition of urban areas. Zipf's law reflects the self-organized nature of urban systems and has been found in most countries \citep{auerbach1913gesetz, jiang2011zipf, rosen1980size, soo2005zipf, zipf1949human}. One expression of Zipf's law is that the probability of a city $i$ larger than size $S$ is inversely proportional to $S$:
\begin{equation}
	P(S_i>S)=kS^{-\alpha}
\end{equation}
where $k$ is a constant and $\alpha=1$. We find that the size distributions of urban clusters in China all follow a power law of $\alpha$ close to $1$ with standard errors less than $0.06$, indicating that Zipf's law holds well for the urban areas delimited by our method (Fig. S3).

\subsection*{Robustness check}
To check the robustness of our method, we expand the analysis to 28 countries and present the results of France (Fig. \ref{fig:pcca_France}) and India (Fig. \ref{fig:pcca_India}) as examples. In the Appendix, we show the results of the remaining countries. In most countries, we obtain similar results as in China: (1) A giant spanning cluster quickly forms when the threshold reaches the critical point. (2) The distribution entropy reaches the maximum around the critical point. (3) The spatial distributions of urban areas delineated by three datasets are similar. In France, the largest cluster is the Paris metropolitan area, the political and economic capital of France. Other larger clusters also correspond to those well-developed regions, such as Marseille, Lyon, and Toulouse (Fig. \ref{fig:pcca_France}). In India, our method can also capture those important urban clusters, such as New Delhi, Mumbai, and Bengaluru (Fig. \ref{fig:pcca_India}). The critical points are 100 population per $km^2$, 6 $km$ per $km^2$, and 1.0 DN value in France, and 1800 population per $km^2$, 2 $km$ per $km^2$, and 2.0 DN value in India. These findings further validate the robustness and generalization of our method.

\begin{figure*}[htbp]
	\centering
	\includegraphics[width=0.8\linewidth]{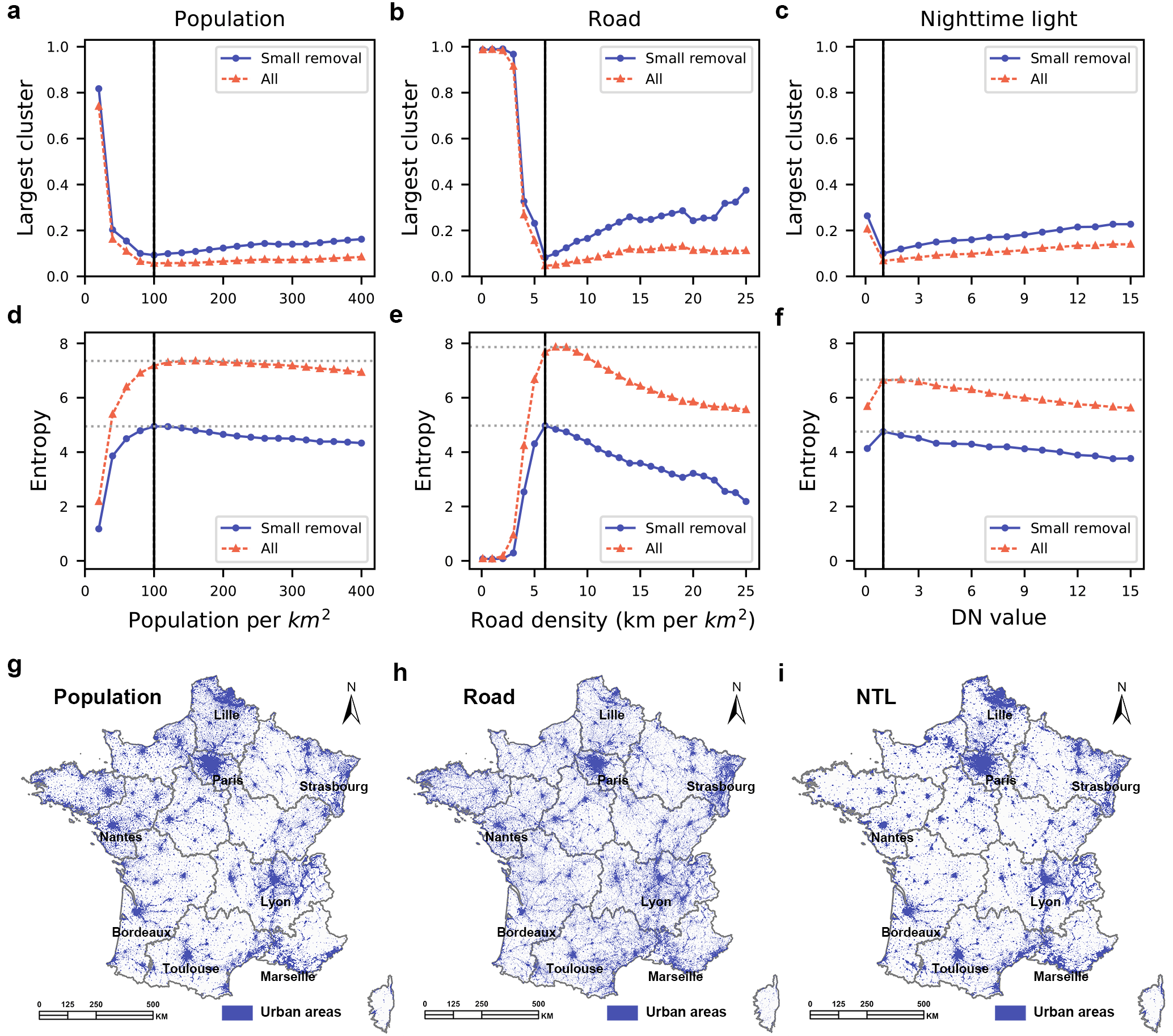}
	\caption{PCCA in France. The largest cluster size for different thresholds in the datasets of population (a), road (b), and NTL (c). The distribution entropy in the datasets of population (d), road (e), and NTL (f). The delineated urban areas in the datasets of population (g), road (h), and NTL (i).}
	\label{fig:pcca_France}
\end{figure*}

\begin{figure*}[htbp]
	\centering
	\includegraphics[width=0.8\linewidth]{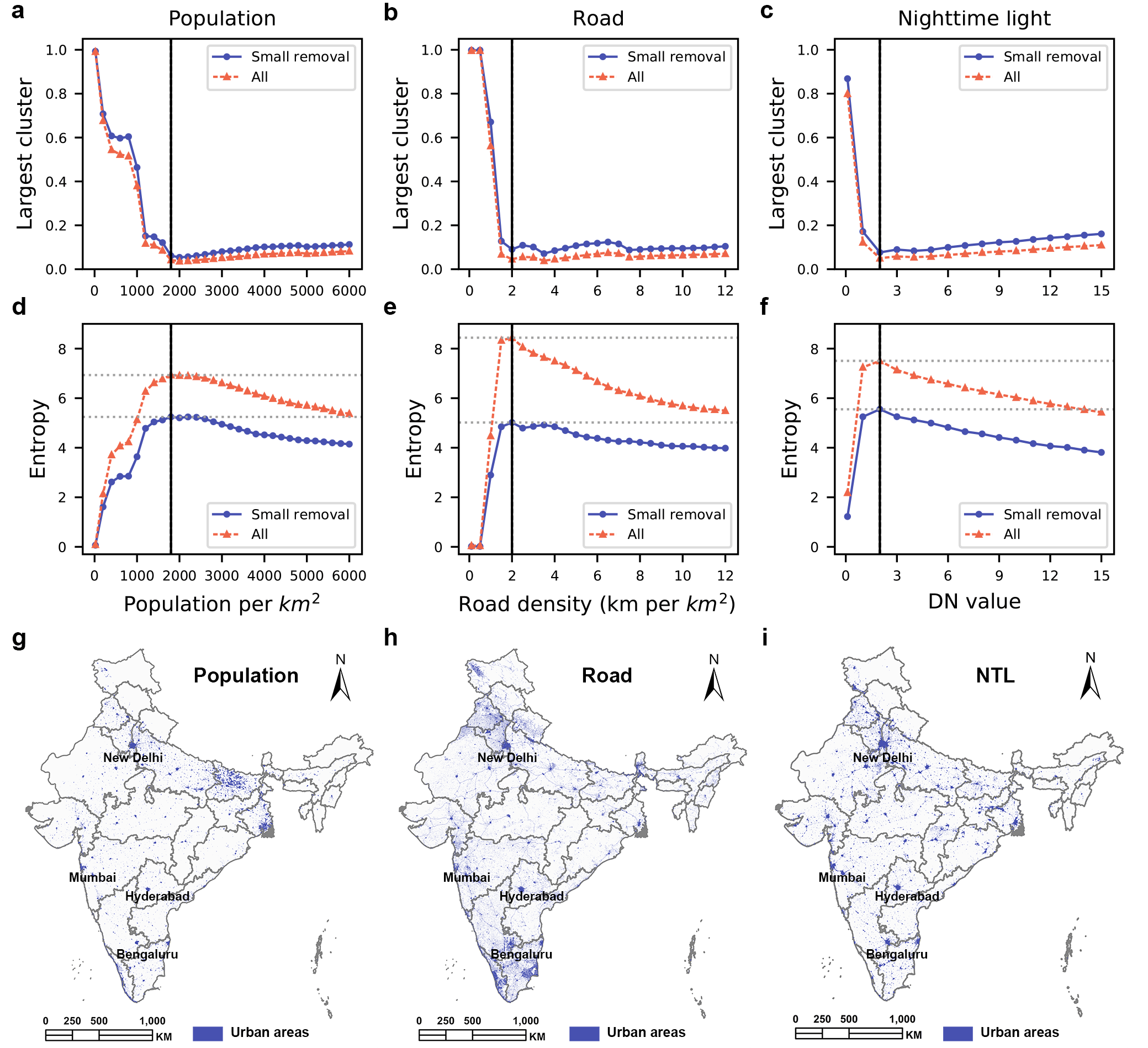}
	\caption{PCCA in India. Same as Fig. \ref{fig:pcca_France}.}
	\label{fig:pcca_India}
\end{figure*}

However, in some countries, we fail to observe a continuous phase transition (entropy is the maximum at the minimum threshold), especially in population and nighttime light datasets. There may be two reasons. On the one hand, limited to resolution, these urban data cannot detect lower urban intensity than the minimum threshold, which mainly occurs in some underdeveloped regions, such as Kazakhstan (Fig. S19) and Chad (Fig. S20). On the other hand, geographical barriers could block the geographical proximity. For example, in Australia, almost all population clusters are distributed in isolated areas along the coast due to the complex geography (e.g., deserts, mountains, rainforests). These geographical barriers break the weak inter-city connections (e.g., population distribution) even at the minimum threshold, while road networks can still connect the cities. Thus, in Australia (Fig. S5), we find the continuous phase transition only in the road dataset.

\subsection*{Accuracy assessment}
To evaluate the accuracy of our results, we compare the derived urban areas with the Landsat-based urban reference maps in 10 cities of 5 countries, including Beijing, Shenzhen, and Taiyuan in China, Ahmedabad and Guwahati in India, Marseille and Toulon in France, Mexico City and San Luis Potos\'{i} in Mexico, and Los Angeles in the United States. These cities span different continents and vary greatly in population size (from 0.2 million to 20 million inhabitants), including both coastal (e.g., Marseille) and inland (e.g., Beijing) cities, as well as cities in developed (e.g., Los Angeles) and developing (e.g., Ahmedabad) countries, which helps test the generality of our method. We use the administrative boundaries collected from GADM (see Data section) to determine the extent of each city. For each test city, we create a reference map with the same spatial resolution of our results and then label each pixel as urban or non-urban by manually interpreting the corresponding area in the Landsat 8 images (available at \url{https://earthexplorer.usgs.gov}). We also use the high-resolution Google Earth imagery to assist the interpretation when it is difficult to distinguish the land use type with Landsat 8 imagery. Our labeled dataset can be accessed through \url{https://github.com/caowenpu56/PCCA}. Besides, we also perform the accuracy analysis on the 2015 Global Human Settlement Layer (GHSL; \cite{corbane2019automated}), one well-known global urban map, to compare our results with GHSL's performance. The GHSL built-up 250m dataset measures the global built-up area density. We downsample the raw GHSL data and convert the density data to the urban/non-urban binary data. Based on previous works \citep{zhou2014cluster-based, zhou2018global}, areas with the built-up density larger than 20\% are labeled as urban areas in GHSL. 

\begin{table*}
	\caption{Accuracy assessment of the derived urban areas by PCCA}
	\begin{ruledtabular}
		\begin{tabular}{p{0.095\linewidth}lllp{0.05\linewidth}lllllllllllllllllllllllll}
			City & Country & Total & Urban & Non-urban & \multicolumn{3}{c}{Population} && \multicolumn{3}{c}{Road} && \multicolumn{3}{c}{NTL} && \multicolumn{3}{c}{Fusion} && \multicolumn{3}{c}{GHSL} \\ \cline{6-8} \cline {10-12} \cline {14-16} \cline {18-20} \cline {22-24}
			& & pixels & pixels & pixels & PA & UA & $\kappa$ && PA & UA & $\kappa$ && PA & UA & $\kappa$ && PA & UA & $\kappa$ && PA & UA & $\kappa$ \\
			\colrule
			Beijing & China & 25,619 & 5,328 & 20,291 & $0.83$ & $0.75$ & $0.73$ && $0.89$ & $0.64$ & $0.67$ && $0.81$ & $0.69$ & $0.67$ && $0.90$ & $0.73$ & \textbf{0.74} && $0.86$ & $0.71$ & $0.71$ \\
			Shenzhen & China & 2,756 & 1,556 & 1,200 & $0.86$ & $0.94$ & \textbf{0.78} && $0.92$ & $0.87$ & $0.76$ && $0.99$ & $0.71$ & $0.51$ && $0.94$ & $0.88$ & \textbf{0.78} && $0.81$ & $0.95$ & $0.75$ \\
			Taiyuan & China & 10,706 & 976 & 9,730 & $0.81$ & $0.76$ & $0.76$ && $0.73$ & $0.73$ & $0.70$ && $0.74$ & $0.63$ & $0.65$ && $0.79$ & $0.78$ & \textbf{0.77} && $0.80$ & $0.72$ & $0.73$ \\
			Ahmadabad & India & 10,398 & 837 & 9,561 & $0.52$ & $0.92$ & $0.64$ && $0.68$ & $0.61$ & $0.61$ && $0.87$ & $0.57$ & $0.65$ && $0.71$ & $0.77$ & \textbf{0.71} && $0.47$ & $0.97$ & $0.61$ \\
			Guwahati & India & 1,152 & 266 & 886 & $0.73$ & $0.93$ & $0.78$ && $0.88$ & $0.79$ & $0.78$ && $0.92$ & $0.78$ & $0.80$ && $0.89$ & $0.88$ & \textbf{0.85} && $0.67$ & $0.94$ & $0.73$ \\
			Marseille & France & 1,229 & 585 & 644 & $0.97$ & $0.76$ & $0.69$ && $0.83$ & $0.69$ & $0.48$ && $0.99$ & $0.73$ & $0.65$ && $0.98$ & $0.76$ & $0.69$ && $0.83$ & $0.89$ & \textbf{0.74} \\
			Toulon & France & 2,368 & 750 & 1,618 & $0.93$ & $0.77$ & $0.76$ && $0.78$ & $0.64$ & $0.55$ && $0.91$ & $0.73$ & $0.71$ && $0.94$ & $0.77$ & \textbf{0.77} && $0.77$ & $0.90$ & $0.76$ \\
			Mexico City & Mexico & 2,107 & 909 & 1,198 & $0.99$ & $0.87$ & \textbf{0.86} && $0.99$ & $0.74$ & $0.70$ && $1.00$ & $0.75$ & $0.72$ && $1.00$ & $0.81$ & $0.80$ && $0.83$ & $0.88$ & $0.74$ \\
			San Luis \newline Potos\'{i} & Mexico & 1,735 & 312 & 1,423 & $0.67$ & $0.94$ & $0.75$ && $0.83$ & $0.69$ & $0.70$ && $0.83$ & $0.79$ & $0.77$ && $0.83$ & $0.88$ & \textbf{0.82} && $0.65$ & $0.96$ & $0.74$ \\
			Los Angeles & USA & 15,398 & 4,858 & 10,540 & $0.89$ & $0.94$ & $0.88$ && $0.97$ & $0.76$ & $0.77$ && $0.99$ & $0.73$ & $0.75$ && $0.97$ & $0.84$ & $0.85$ && $0.94$ & $0.93$ & \textbf{0.91} \\
			Total & & 73,468 & 16,377 & 57,091 & $0.85$ & $0.84$ & \textbf{0.80} && $0.89$ & $0.71$ & $0.72$ && $0.90$ & $0.70$ & $0.72$ && $0.91$ & $0.79$ & \textbf{0.80} && $0.84$ & $0.83$ & $0.79$ \\
		\end{tabular}
	\end{ruledtabular}
	\footnotetext{Note: Results with the highest $\kappa$ for each city are marked in bold.}
	\label{tab:accuracy}
\end{table*}

For our three results (population, road, and nighttime light) and GHSL data, we compare all pixels located in the test extent of each city with the urban reference maps pixel by pixel, and then calculate the confusion matrices. The number of labeled pixels (urban/non-urban) in the reference maps for each city is listed in Table \ref{tab:accuracy}. Due to the large number of non-urban pixels, the overall accuracy (OA) is high in all cites. The OAs of total test areas are $0.93$, $0.89$, and $0.89$ in the population, road, and nighttime light dataset, respectively. In Table \ref{tab:accuracy}, we also list the user's and producer's accuracy of the urban class, and the Kappa coefficients ($\kappa$) for each city. Our results show good agreement with the urban reference maps, with the $\kappa$ of $0.64-0.88$ (mean: $0.76$, sd: $0.07$) in population dataset, $0.48-0.78$ (mean: $0.67$, sd: $0.10$) in road dataset, and $0.51-0.80$ (mean: $0.69$, sd: $0.08$) in nighttime light dataset. These Kappa coefficients calculated on a single dataset are similar to the GHSL data ($\kappa=0.61-0.91$, mean: $0.74$, sd: $0.07$). However, we highlight that our PCCA method is based on a physical model (i.e., percolation) and only has one free parameter (i.e., threshold) that can be derived through the percolation process. These properties make our method better in interpretation and efficiency.

\begin{figure*}
	\centering
	\includegraphics[width=0.8\linewidth]{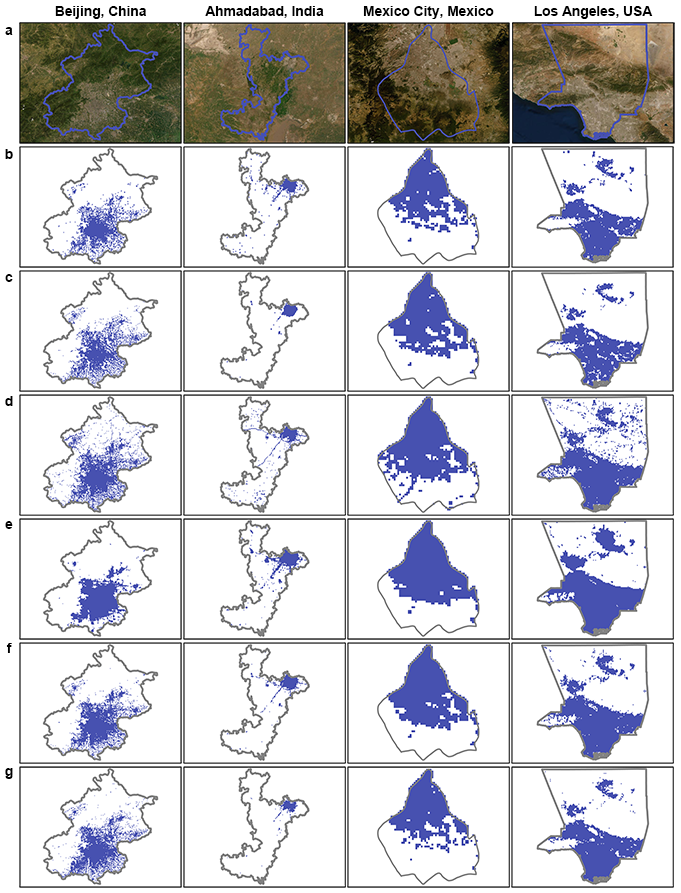}
	\caption{A comparison of the derived urban areas by population (c), road (d), and nighttime light (e) dataset, the fusion results (f), the GHSL data (e) with the Landsat-based urban reference maps (b) in Beijing, Ahmadabad, Mexico City, and Los Angeles. (a) Google Earth imagery.}
	\label{fig:validation_cities}
\end{figure*}

Additionally, since urban areas are derived from multi-source datasets, we can further generate a fused urban area map to improve the results. Specifically, we merge the urban areas delimited by three datasets; and extract urban areas identified in at least two datasets. This process is similar to the majority voting in the ensemble model, which can improve the accuracy and stability of the results in different regions \citep{trianni2015scaling}. Fig. \ref{fig:validation_cities} presents the urban maps of the fusion results, the GHSL data, and the reference data. Visually, our method captures the accurate urban extent for all test cities, and the fusion results match well with the reference data. Statistically, the accuracy scores indicate that the fusion results have improved accuracy and stability over the results from a single dataset. The mean $\kappa$ increases from $0.67-0.76$ to $0.78$, while the standard deviation of $\kappa$ decreases from $0.07-0.10$ to $0.05$ (Table \ref{tab:accuracy}). Besides, the fusion results perform better than the GHSL data with a larger $\kappa$ in 8/10 cities. We also apply a student's t-test to the $\kappa$ between the fusion results and the GHSL data in test cities, and the results show that the improvement is statistically significant ($p$-value $<0.05$). 

\subsection*{Optimal threshold and socio-economic development}
Table \ref{tab:pcca_world} presents the optimal threshold ($D_{pop}$, $D_{road}$, $D_{ntl}$), entropy at the threshold ($E_{pop}$, $E_{road}$, $E_{ntl}$), and Zipf exponent ($\alpha_{pop}$, $\alpha_{road}$, $\alpha_{ntl}$) for each country and each dataset. We find that $D_{pop}$ varies greatly from country to country, while $D_{road}$ and $D_{ntl}$ change less since the development of road networks and nighttime light is limited by geospatial space. For each country, the entropies at the critical point in three datasets are similar, indicating the similar size distributions of the delineated urban areas. Meanwhile, the Zipf exponents of size distributions fall within $[0.75,1.25]$ in 24/28 (population), 20/24 (road), and 24/28 (nighttime light) countries, which means that Zipf's law holds well in most countries.

\begin{table*}
	\caption{PCCA in 28 countries}
	\begin{ruledtabular}
		\begin{tabular}{lllllllccc}
			Country & $D_{pop}$ & $E_{pop}$ & $D_{road}$ & $E_{road}$ & $D_{ntl}$& $E_{ntl}$ & $\alpha_{pop}$ & $\alpha_{road}$ & $\alpha_{ntl}$\\ 
			\colrule
			Algeria & $300$ & $4.18$ & $2.5$ & $3.83$ & $2.0$ & $4.42$ & $0.97\pm 0.05$ & $1.12\pm 0.06$ & $1.02\pm 0.05$ \\
			Argentina & $20^*$ & $5.05$ & $2.0$ & $4.90$ & $0.1^*$ & $5.36$ & $1.07\pm 0.04$ & $1.25\pm 0.04$ & $1.26\pm 0.05$ \\
			Australia & $20^*$ & $4.11$ & $2.0$ & $4.18$ & $0.1^*$ & $4.67$ & $1.05\pm 0.03$ & $1.29\pm 0.03$ & $1.13\pm 0.05$ \\
			Bolivia & $40$ & $3.38$ & $1.4$ & $4.51$ & $0.1^*$ & $3.25$ & $0.82\pm 0.05$ & $1.36\pm 0.05$ & $1.10\pm 0.09$ \\
			Brazil & $20^*$ & $5.92$ & $1.5$ & $5.58$ & $0.1^*$ & $5.93$ & $1.24\pm 0.03$ & $1.15\pm 0.03$ & $1.22\pm 0.03$ \\
			Canada & $20^*$ & $4.57$ & $-$ & $-$ & $0.5$ & $4.85$ & $1.06\pm 0.02$ & $-$ & $1.05\pm 0.01$ \\
			Chad & $140$ & $3.38$ & $0.8$ & $4.64$ & $0.1^*$ & $2.15$ & $0.98\pm 0.14$ & $1.20\pm 0.05$ & $0.78\pm 0.11$ \\
			China & $700$ & $6.49$ & $2.5$ & $5.80$ & $3.0$ & $5.34$ & $1.21\pm 0.04$ & $1.15\pm 0.02$ & $1.00\pm 0.03$ \\
			Colombia & $80$ & $4.07$ & $1.6$ & $3.70$ & $0.1^*$ & $4.44$ & $0.81\pm 0.05$ & $0.87\pm 0.09$ & $1.08\pm 0.05$ \\
			D. R. Congo & $100$ & $4.74$ & $1.0$ & $5.70$ & $0.1^*$ & $3.12$ & $1.07\pm 0.06$ & $1.26\pm 0.05$ & $1.37\pm 0.25$ \\
			France & $100$ & $4.95$ & $6.0$ & $4.97$ & $1.0$ & $4.75$ & $1.11\pm 0.02$ & $1.07\pm 0.02$ & $1.02\pm 0.02$ \\
			Germany & $200$ & $4.12$ & $11.0$ & $4.23$ & $0.5$ & $4.34$ & $1.10\pm 0.04$ & $1.12\pm 0.03$ & $1.13\pm 0.03$ \\
			Honduras & $80$ & $3.10$ & $-$ & $-$ & $0.1^*$ & $3.54$ & $0.82\pm 0.07$ & $-$ & $1.04\pm 0.08$ \\
			India & $1800$ & $5.24$ & $2.0$ & $5.02$ & $2.0$ & $5.55$ & $0.98\pm 0.02$ & $1.19\pm 0.02$ & $1.05\pm 0.03$ \\
			Indonesia & $1600$ & $3.39$ & $2.5$ & $4.85$ & $1.0$ & $3.78$ & $0.89\pm 0.09$ & $1.03\pm 0.04$ & $1.00\pm 0.06$ \\
			Kazakhstan & $20^*$ & $4.15$ & $1.0$ & $5.78$ & $0.1^*$ & $4.79$ & $0.77\pm 0.05$ & $1.29\pm 0.04$ & $1.11\pm 0.04$ \\
			Libya & $140$ & $2.05$ & $0.8$ & $3.76$ & $0.1^*$ & $2.98$ & $0.94\pm 0.08$ & $1.24\pm 0.07$ & $1.01\pm 0.12$ \\
			Mexico & $300$ & $4.59$ & $1.8$ & $5.15$ & $3.0$ & $4.35$ & $1.06\pm 0.05$ & $1.00\pm 0.06$ & $0.87\pm 0.05$ \\
			New Zealand & $20^*$ & $3.08$ & $1.4$ & $4.15$ & $0.1^*$ & $3.62$ & $0.99\pm 0.05$ & $1.17\pm 0.03$ & $1.13\pm 0.12$ \\
			Nicaragua & $100$ & $2.47$ & $-$ & $-$ & $0.1^*$ & $2.25$ & $0.72\pm 0.06$ & $-$ & $1.18\pm 0.17$ \\
			Papua New Cuinea & $320$ & $1.96$ & $0.8$ & $4.71$ & $0.1^*$ & $2.49$ & $0.39\pm 0.08$ & $0.96\pm 0.05$ & $1.54\pm 0.37$ \\
			Peru & $40$ & $4.63$ & $1.4$ & $5.02$ & $0.1^*$ & $4.46$ & $0.73\pm 0.03$ & $1.19\pm 0.06$ & $0.96\pm 0.05$ \\
			Saudi Arabia & $40$ & $3.48$ & $-$ & $-$ & $1.5$ & $4.17$ & $0.74\pm 0.03$ & $-$ & $0.82\pm 0.03$ \\
			Spain & $60$ & $4.02$ & $5.5$ & $3.85$ & $0.5$ & $4.36$ & $1.13\pm 0.03$ & $1.12\pm 0.04$ & $0.95\pm 0.04$ \\
			Sudan & $20^*$ & $2.19$ & $1.0$ & $4.58$ & $0.1^*$ & $3.40$ & $0.79\pm 0.07$ & $1.24\pm 0.06$ & $0.87\pm 0.06$ \\
			Sweden & $20^*$ & $3.95$ & $3.5$ & $4.05$ & $0.1^*$ & $4.86$ & $1.04\pm 0.04$ & $1.09\pm 0.05$ & $1.13\pm 0.03$ \\
			Ukraine & $60$ & $5.51$ & $2.5$ & $5.25$ & $0.1^*$ & $5.13$ & $1.17\pm 0.04$ & $1.22\pm 0.03$ & $1.27\pm 0.03$ \\
			USA & $300$ & $5.62$ & $4.0$ & $6.00$ & $1.0$ & $5.94$ & $0.96\pm 0.02$ & $1.11\pm 0.01$ & $0.92\pm 0.01$ \\
		\end{tabular}
	\end{ruledtabular}
	\footnotetext{$^*$: no continuous phase transition and the minimum threshold is used.}
	\footnotetext{$-$: no available data.}
	\footnotetext{$\pm$: standard error.}
	\label{tab:pcca_world}
\end{table*}

\begin{figure*}
	\centering
	\includegraphics[width=0.8\linewidth]{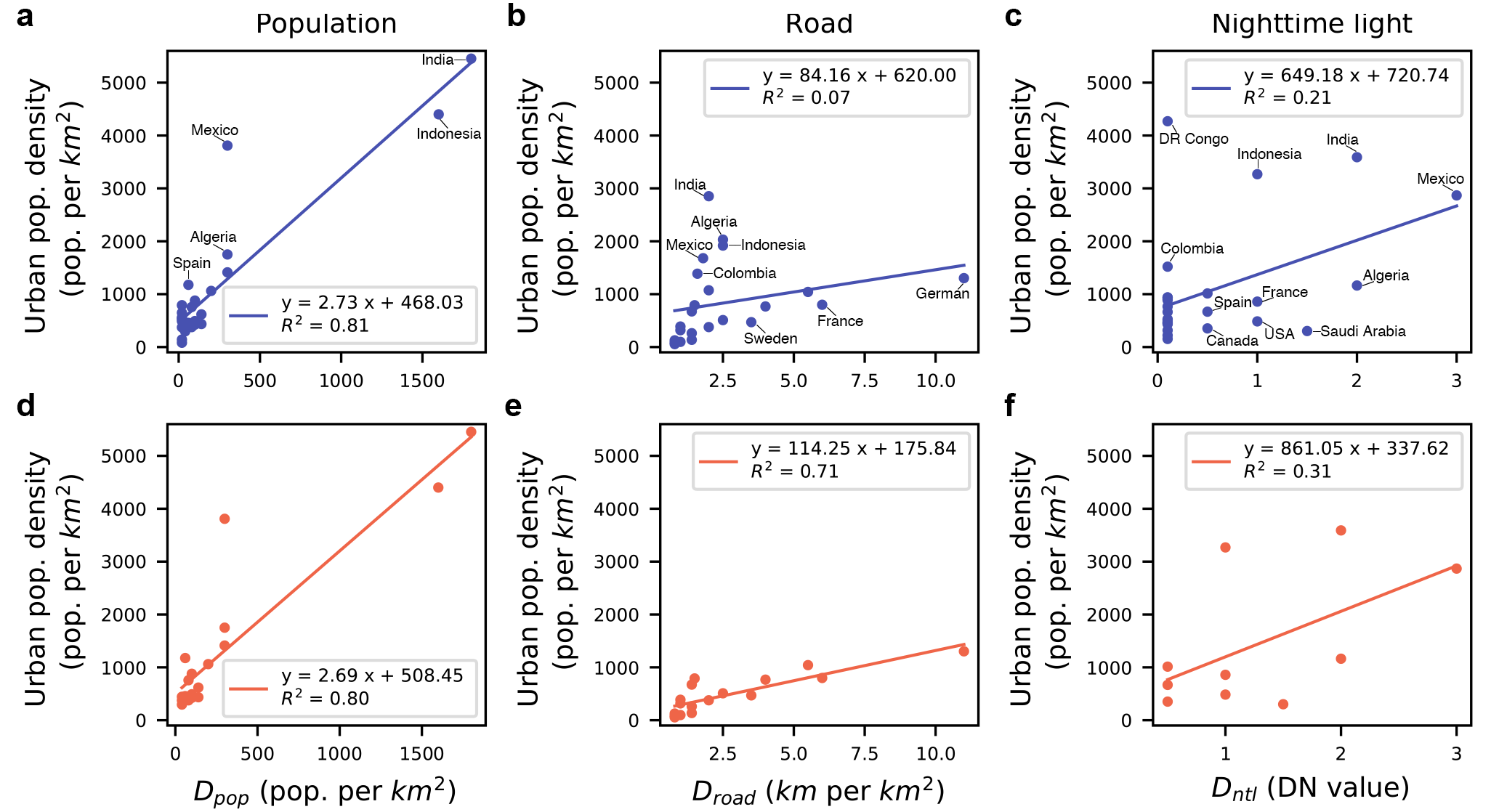}
	\caption{Relationship between the optimal threshold and urban population density. All countries are included in the datasets of population (a), road (b), and NTL (c). Some countries are removed in the datasets of population (d), road (e), and NTL (f). In population and NTL dataset, we remove the countries without a continuous phase transition; in the road dataset, we remove those with large population size and poor transportation infrastructure (India, Algeria, Indonesia, Mexico, Colombia, and Argentina).} 
	\label{fig:socio-economic}
\end{figure*}

Furthermore, we explore the relationship between the optimal thresholds and countries' socio-economic indicators. Here, we use the urban population density as the proxy for socio-economic development. We calculate the urban population density by dividing the total population (WorldPop data) that fall within the urban clusters for each country. Intuitively, the population threshold $D_{pop}$ is highly correlated with urban population density (the $R^2$ is 0.81, Fig. \ref{fig:socio-economic}a), and $D_{pop}$ of each country is about $1/3$ of the country's urban population density. However, thresholds of road ($D_{road}$) and nighttime light ($D_{ntl}$) have weak positive correlations with urban population density (Fig. \ref{fig:socio-economic}b,c). This may result from the country's slow development of transportation infrastructure, which mainly occurs in some developing countries with large population size, such as India, Algeria, and Mexico. Besides, the OSM data quality varies greatly across different countries, and many road lines are not captured by OSM in some developing countries. Therefore, $D_{road}$ is smaller than the actual urban road density threshold in some countries. (The $R^2$ between $D_{road}$ and urban population density becomes 0.71 if removing some special cases, as shown in Fig. \ref{fig:socio-economic}e.)

\begin{figure*}
	\centering
	\includegraphics[width=0.8\linewidth]{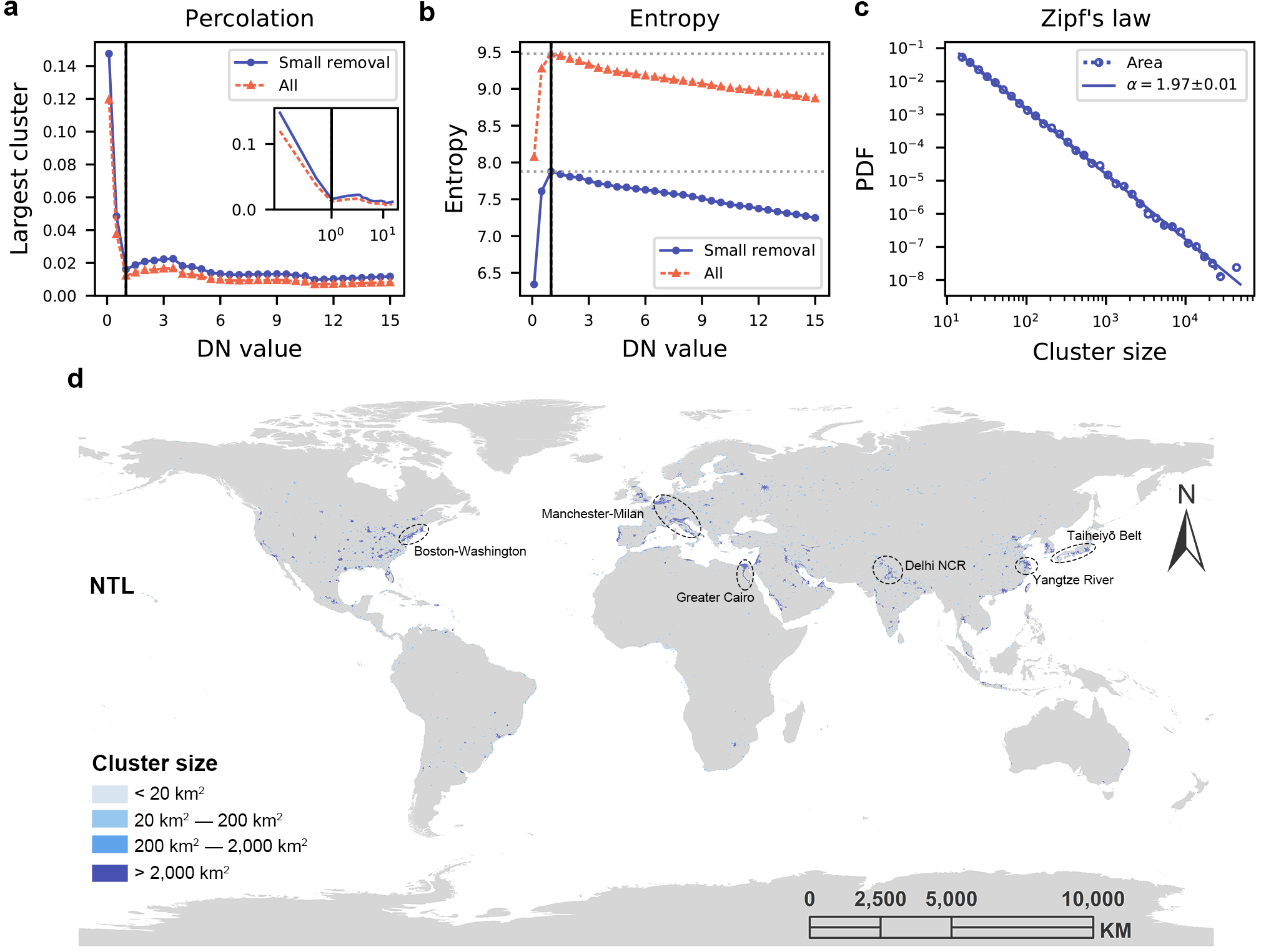}
	\caption{PCCA applied to the world using nighttime light data. (a) Largest cluster sizes. (b) Distribution entropies. (c) PDF and fitting line of urban cluster areas (Zipf's law). (d) Urban areas by PCCA.}
	\label{fig:pcca_World}
\end{figure*}

Finally, we apply the method to the entire world using nighttime light data and present the delineated urban areas in Fig. \ref{fig:pcca_World}. Similar to the country level findings, the largest cluster size goes through a critical point, and the maximum distribution entropy is exactly at this point. Then, we obtain the optimal threshold -- 1.0 DN value. Through this world map (Fig. \ref{fig:pcca_World}), we find that the urbanization level is much higher in North America, Europe, and East Asia. The top six urban clusters correspond to the Manchester-Milan (Europe), the Greater Cairo (Egypt), the Yangtze River Delta (China), the Boston-Washington (USA), the Delhi National Capital Region (India), and the Taiheiy\a=o Belt (Japan) megalopolises (Fig. \ref{fig:pcca_World}). The size distribution of urban areas of the world also fits well with Zipf's law, with an exponent of $0.97\pm 0.01$. We note that different dimensions of urban areas are captured at different spatial scales. At the country level (such as Fig. \ref{fig:natural-cities_China}), we delimit the metropolitan areas; while at the world level (Fig. \ref{fig:pcca_World}), due to the differences in the economic basis of each country, we delimit those large mega-regions or urban corridors \citep{georg2016new}. For example, the development level of the capitals of some African countries is even far less than that of rural areas in some developed countries. Therefore, the meanings of our extractions are different at different spatial level, and they depend on applications.

\section*{Discussion and conclusions}

In summary, we propose a `percolation-based city clustering algorithm' to extract urban areas from multi-source urban data (nighttime light images, population data, and road networks). Our method only needs one parameter (urban/non-urban threshold), which can be derived solely through the input data themselves by considering the critical nature of urban systems. The derived urban areas are validated in several parts of the world, and they can be further improved by data fusion.

The contributions of this study can be summarized in three aspects. First, we bridge the gap between remote sensing and emerging urban data in the task of delimiting urban areas. Our study has demonstrated that despite great differences, different urban data can reflect the similar socio-economic dynamics of cities. Second, our method provides a consistent measurement of urban areas since the optimal threshold is derived automatically and under the same criteria. With our method, urban development can be measured under a unified standard, which allows comparisons across different countries and periods. Third, we show the potential of open-source data in delimiting urban areas. With the proposed method, we can produce reliable urban area maps from these publicly available data, which is especially helpful for those developing regions with limited survey data. Our study is also an attempt for applying complexity science to solve traditional urban problems and could deepen our understanding of urban systems.

There are still some limitations in this study, and several improvements can be explored in future work. First, due to the limited availability of temporal urban data, we have not been able to track the changes of urban areas over time. Such analysis could be possible with more detailed spatio-temporal data in the future. Second, it is meaningful to study the factors that influence the values of optimal thresholds. Possible explanations can be complicated for geographical, social, or economic reasons. For example, environmental awareness can cause a decrease in brightness of nighttime light in some regions of Europe \citep{bennie2015contrasting}. Third, the differences of the urban areas delineated by multi-source data are also worth exploring, as they reflect the inconsistent configuration of urban elements. However, how to explain these differences is full of challenges and requires more in-depth study.

\section*{Data availability}
All open-source datasets are available through the websites described in the data section. The PCCA codes and the maps of the delineated urban areas can be obtained through \url{https://github.com/caowenpu56/PCCA}.

\section*{Acknowledgements}
This research was supported by the National Natural Science Foundation of China (Grant Nos.41625003, 41801299, 41830645) and the China Postdoctoral Science Foundation (2018M630026).

\bibliography{PCCA_arXiv.bib}

\end{document}